\documentstyle[fleqn,epsf,11pt]{article}

\catcode`@=11
\def\twoup{\twocolumn\sloppy\flushbottom\parindent 2em
    \parskip .33\baselineskip \leftmargini 2em\leftmarginv
.5em\leftmarginvi .5em \oddsidemargin 0in \evensidemargin 0in
\columnsep .4in \footheight 0pt \textwidth 10in \topmargin -.4in
\headheight 0pt \topskip 0in \textheight 6.9in \footskip 30pt \hoffset
-.5in \voffset -.25in
\def\@oddfoot{\hfil\thepage\hfil\addtocounter{page}{1}
\hspace{\columnsep}\hfil\thepage\hfil} \let\@evenfoot\@oddfoot
\def\@oddhead{} \def\@evenhead{} \newcommand{\figsize}{4.5in}
\typeout{DEFAULT ORIENTATION: LANDSCAPE - DEFAULT PAPER SIZE: US
LEGAL}}

\def\twoupA4{\twocolumn\sloppy\flushbottom\parindent 2em
    \parskip .33\baselineskip \leftmargini 2em\leftmarginv
.5em\leftmarginvi .5em \oddsidemargin 0in \evensidemargin 0in
\columnsep .4in \footheight 0pt \textwidth 10.3in \topmargin -.4in
\headheight 0pt \topskip 0in \textheight 6.7in \footskip 30pt \hoffset
-.5in \voffset -.25in
\def\@oddfoot{\hfil\thepage\hfil\addtocounter{page}{1}
\hspace{\columnsep}\hfil\thepage\hfil} \let\@evenfoot\@oddfoot
\def\@oddhead{} \def\@evenhead{} \newcommand{\figsize}{4.7in}
\typeout{DEFAULT ORIENTATION: LANDSCAPE - DEFAULT PAPER SIZE: DIN A4}}
\catcode`@=12

\def\A4{\setlength{\oddsidemargin}{0in}
    \setlength{\textwidth}{6.3in} \setlength{\textheight}{8.5in}
\setlength{\topmargin}{0in} \renewcommand{\baselinestretch}{1.25}
\renewcommand{\arraystretch}{.8} \newcommand{\figsize}{6.3in}
\typeout{DEFAULT ORIENTATION: PORTRAIT - DEFAULT PAPER SIZE: DIN A4}}

\twoup     

\newcommand{\tri}{{\cal T}}
\newfont{\sm}{cmr8}
\newcommand{\mathrm}[1] {\mbox{\sm #1}}
\begin{document}
\protect\pagestyle{empty}
\title{Critical Slowing Down of Cluster Algorithms for Ising Models
Coupled to \mbox{2--d} Gravity}
\author{Mark Bowick, Marco Falcioni,\\ Geoffrey Harris and Enzo
Marinari$^{(*)}$\\[.8em] Dept. of Physics and NPAC,\\Syracuse
University,\\ Syracuse, New York, 13244--1130, USA\\[.8em]
{\footnotesize {\tt bowick falcioni gharris@npac.syr.edu
marinari@roma1.infn.it}}\\[.8em] {\small (*) and Dipartimento di
Fisica and INFN,}\\ {\small Universit\`a di Roma ``Tor Vergata'',}\\
{\small Viale Della Ricerca Scientifica, 00133 Roma, Italy}}
\date{\today}

\maketitle
\begin{quote}
\begin{center} {\bf Abstract} \end{center} \medskip
{\small We simulate single and multiple Ising models coupled to
\mbox{2--d} gravity using both the Swendsen-Wang and Wolff algorithms
to update the spins.  We study the integrated autocorrelation time and
find that there is considerable critical slowing down, particularly in
the magnetization.  We argue that this is primarily due to the local
nature of the dynamical triangulation algorithm and to the generation
of a distribution of baby universes which inhibits cluster growth.}
\end{quote}
\vfill
\begin{flushright} {\parbox{1.8in}{SU--HEP--93--4241--560\\SCCS 568\\
{\bf \tt hep-lat/9311036}}}
\end{flushright}
\newpage
\pagestyle{plain}
\setcounter{page}{1}

Considerable work has been devoted to the study of the performance of
cluster algorithms in reducing the critical slowing down of many
statistical models.  The Swendsen-Wang \cite{SW} and the Wolff
\cite{WOLFF} algorithms have proven very effective in beating the
critical slowing down (CSD) exhibited by these models when simulated
with a standard Metropolis \cite{METRO} update. Reviews on cluster
algorithms and CSD may be found on refs.~\cite{REVIEWS}. In recent
years, the numerical study of Potts models coupled to \mbox{2--d}
gravity has received much attention \cite{DTRS1,DTRS2,DTRS3}, aided as
well by increased analytical understanding of these models
\cite{ANALYTICAL}.  Cluster algorithms have proved useful in saving
computational effort on the update of the Potts variables.  The
present study is motivated by the fact that there is little
understanding of the actual extent of the improvement achieved in
these simulations.  It is worthwhile, therefore, to measure CSD in the
case of simple Ising spins coupled to a dynamical lattice in order to
quantify the performance of cluster algorithms.  We find that there is
considerable CSD, especially in the magnetization, and we relate this
to the dynamics of cluster formation on a random lattice.

We shall consider a model in which $n_{\mathrm{s}}$ Ising spins are
attached to the vertices of triangulations.  The triangulations are
characterized by their adjacency matrix $C_{\mathrm{ij}}$, which
equals $1$ if i and j are neighbors and vanishes otherwise.
$C_{\mathrm{ij}}$ is the discrete analogue of the world-sheet metric
$g_{\mathrm{ij}}$.  We shall restrict ourselves to the set of
triangulations with $N$ vertices $\tri_N$ containing only loops of
length $3$ or greater and vertices of coordination number of at least
$3$.  The triangulation has a fixed toroidal topology.  We simulate a
theory determined by the partition function
\begin{equation}
Z_N = \sum_{T \in \tri_N} \sum_{\sigma_{\mathrm{i} = \pm 1}}
\exp \left( -\beta
\sum_{\alpha=1}^{n_{\mathrm{s}}} \sum_{\mathrm{{i,j=1}}}
^{N} C_{\mathrm{ij}}(T) \sigma_{\mathrm{i}}^{\alpha}
\sigma_{\mathrm{j}}^{\alpha}\right),
\protect\label{partfn}
\end{equation}
where $\alpha$ labels the spin species.  In refs.
\cite{TWOISING,GEOFF} we investigated in detail this model for the
cases $n_{\mathrm{s}} =$ 1 and 2. We measured spin susceptibility and
percolation critical exponents using finite-size scaling and showed
that logarithmic corrections to scaling were essential for agreement
between the measured and theoretical exponents.  In this paper, we
deal with the cases $n_{\mathrm{s}} =$ 1, 2 and 3, concentrating on
the issue of CSD and its origin.

The standard way of implementing the partition function (\ref{partfn})
via a Monte Carlo simulation is to use the Swendsen-Wang (SW) or Wolff
cluster algorithm to update the spin variables and to use the ``link
flip'' \cite{FLIP} to simulate the sum over all triangulations.  To
implement a SW update one first divides all of the spins into
Fortuin-Kasteleyn (FK) clusters \cite{FK}.  These clusters of bonded
spins are created by introducing bonds between same sign spins with
probability $p=1-\exp(-2\beta)$.  Then, one flips all clusters with
probability one-half.  The Wolff algorithm consists of randomly
choosing a spin, constructing a FK cluster around it and flipping the
cluster with probability one.  To compare the autocorrelation times of
the SW and Wolff algorithms one should define the Wolff update so that
both algorithms require comparable CPU time.  For this reason we
choose a Wolff update to consist of consecutive flips of FK clusters
that reverse the sign of at least 40\% of the spins.  One alternative
to this would be to scale the correlation times using the average
cluster size \cite{WOLFF2}.  Each spin update precedes a mesh update,
in which we attempt to flip $3N$ randomly chosen links, $N$ being the
number of vertices of the triangulation.  Our implementation ensures
that the relative number of mesh and spin updates is roughly
equivalent.

The observables that we analyzed are the energy density, the
magnetization density, the susceptibility (namely the magnetization
squared) and the average value of $\vert q - 6 \vert$, where $q$ is
the coordination number of a vertex of the triangulation.  In some of
the simulations that employed SW updates, we also measured the mean
size of pure percolation and FK clusters.  The mean FK cluster size
${\cal{S}}_{\rm{FK}}$ is given by the quantities $\langle s
\rangle_{\mathrm{Wolff}}$ and $\langle s^2
\rangle_{\mathrm{SW}}/\langle s \rangle_{\mathrm{SW}}$;  $s$ denotes
the number of sites constituting a cluster and averages are taken over
the distribution of clusters built in the Wolff and SW algorithms
respectively.  The magnetic Ising observables are directly related to
the structure of these clusters \cite{PERCOLBOOK}. In particular, for
$\beta \leq \beta_c$, ${\cal{S}}_{\rm{FK}}$ is equal to the
susceptibility, defined as
\begin{equation}
\chi = {\frac{\beta}{N}}  \langle M^2 \rangle,
\label{suscept}
\end{equation}
where $M$ denotes the average magnetization density.  Actually, this
alternative definition of $\chi$ is used as a reduced variance
estimator \cite{SWEENY}.

Before moving on to present our results, we discuss how we estimated
the autocorrelation times.  It is known \cite{SOKAL} that the
following relation holds between the estimators of the naive and true
variance of an observable $O$;
\begin{equation}
\mbox{Var}(O)_{\mathrm{true}} \simeq 2 ~ \tau_{\mathrm{int}}
{}~\mbox{Var}(O)_{\mathrm{naive}}.
\label{tauint}
\end{equation}
For a subset of our data, we also directly measured the
autocorrelation function and through standard methods \cite{SOKAL}
computed $\tau_{\mathrm{int}}$.  We verified (see also \cite{JANKE})
that both of the above techniques gave consistent values of
$\tau_{\mathrm{int}}$.  The variances of the observables were
extracted from the raw data using the {\sl binning\/} method.  To
extract the dynamic exponent, we fit the data according to the scaling
ansatz
\begin{equation}
\tau_{\mathrm{int}} \propto N^{z/d_H}.
\label{ansatz}
\end{equation}
It is difficult to estimate the linear size of random triangulations.
Therefore, we shall leave explicit the dependence of the exponent upon
the intrinsic Hausdorff dimension $d_H$ of the triangulation and fit
our data using the total area $N$.  The values of $z/d_H$ were
extracted from the auto-correlation data using a log-log regression
fit, excluding from the fit the results for the smaller volumes, since
they are affected the most by finite-size effects.  The values of
$\chi^2$ per degree of freedom were always of order one.  In the case
of the magnetization for the $n_{\mathrm{s}}=2$ and 3 models, Wolff
algorithm, we extracted the exponent using only the two largest
lattices, since there are larger finite-size effects.

\begin{table}[th]
\begin{center}
\begin{tabular}{lccc}\hline
& \multicolumn{3}{l}{$z/d_H$} \\ \cline{2-4} Model & Metropolis & SW &
Wolff \\ \hline $n_{\mathrm{s}} = 1$ & $.85 \pm .06$ & $.58 \pm .05$ &
$.54 \pm .05$
\\ $n_{\mathrm{s}} = 2^{(*)}$ & $.95 \pm .05$ & $.62 \pm .06$ & $.58
\pm .09$ \\ $n_{\mathrm{s}} = 3^{(*)}$ & $.9 \pm .1$ & $.49 \pm .08$ &
$.55 \pm .1$ \\ \hline
\end{tabular}
\protect{\caption{Critical exponent $z/d_H$ for the
Magnetization from fits. (*)~These numbers are not reliable; we
discuss this point in the text.}\label{table1}}
\end{center}
\end{table}

We present results for three different models, $n_{\mathrm{s}}=1$, 2
and 3.  Each model was simulated with the SW, Wolff and Metropolis
algorithms.  The $n_{\mathrm{s}}=1$ model was simulated at the
critical value of $\beta$, which is known analytically
\cite{TWOISING}.  For the other models we chose the $\beta$ value by
looking at the peak of the susceptibility and the intersection of the
Binder's cumulant \cite{BINDER} curves \cite{TWOISING}.  Each
simulation (model and algorithm) was run at four or five values of $N$
(512, 1024, 2048, 4096 and 8192 in the $n_{\mathrm{s}}=1$ model) and
each consisted of $10^5$ thermalization sweeps and 3--5$\times10^5$
measurement sweeps. Measurements were taken every sweep.

\begin{table}[th]
\begin{center}
\begin{tabular}{lccc}\hline
& \multicolumn{3}{l}{$z/d_H$} \\ \cline{2-4} Model & Metropolis & SW &
Wolff \\ \hline $n_{\mathrm{s}} = 1$ & $.62 \pm .03$ & $.057 \pm .005$
& $.04 \pm .03$
\\ $n_{\mathrm{s}} = 2$ & $.35 \pm .1$ & $.08 \pm .02$ & $.17 \pm .08$
\\ $n_{\mathrm{s}} = 3$ & $.5 \pm .1$ & $.05 \pm .04$ & $.37 \pm .08$
\\ \hline
\end{tabular}
\protect{\caption{Critical exponent $z/d_H$ for the Energy from
fits.}\label{table2}}
\end{center}
\end{table}

\begin{figure}[th]
\epsfxsize=\figsize \epsfbox{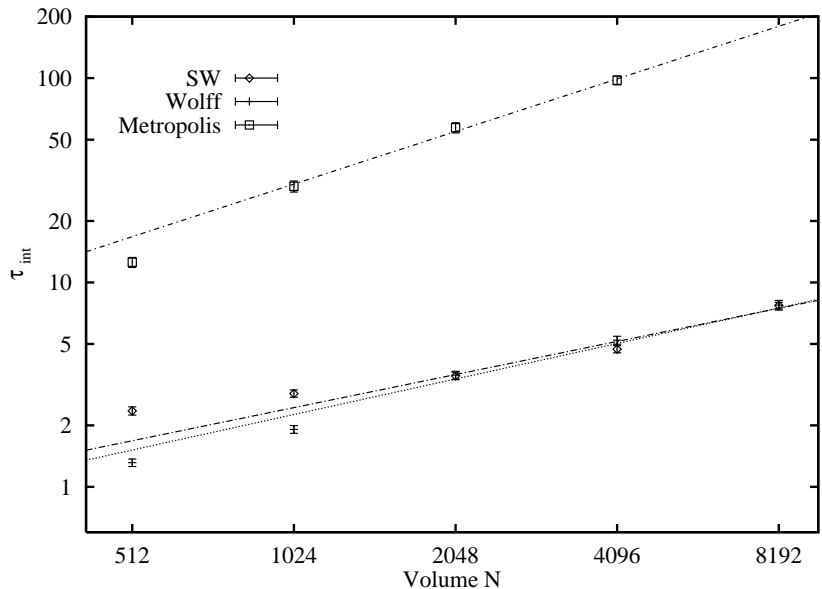}
\protect\caption{\protect\label{fig:1genmag} Comparison of the integrated
autocorrelation times for the magnetization in the $n_{\mathrm{s}}=1$
model.  The dashed lines are log-log regression fits.}
\end{figure}

\begin{figure}[th]
\epsfxsize=\figsize \epsfbox{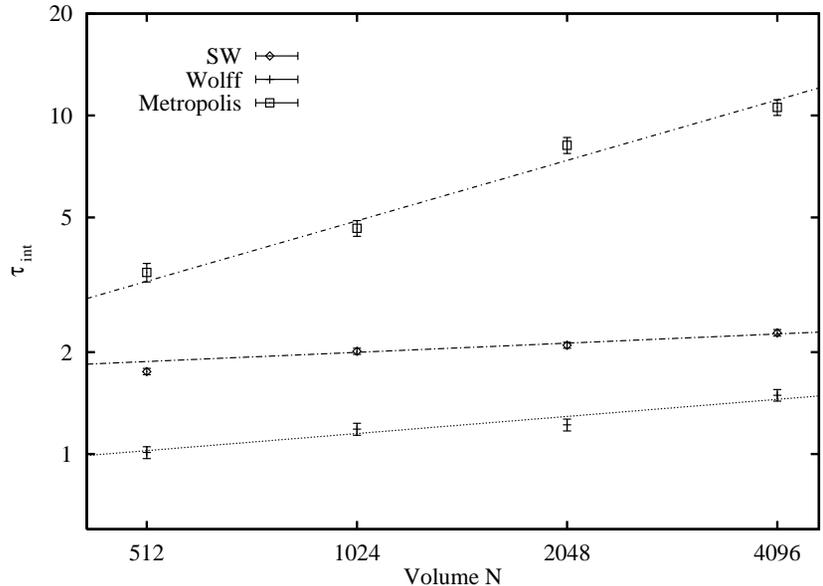}
\protect\caption{\protect\label{fig:2genene} Comparison of the integrated
autocorrelation times for the energy in the $n_{\mathrm{s}}=2$ model.}
\end{figure}

{}From the analysis presented in figures~\ref{fig:1genmag},
\ref{fig:1genene}, \ref{fig:2genmag} and
\ref{fig:2genene} and tables~\ref{table1} and \ref{table2}, we deduce
the following:

\begin{enumerate}
\item There is considerable critical slowing down in these
models.  Figure~\ref{fig:1genmag} shows the improvement gained by the
use of cluster algorithms---the autocorrelation times {\sl and} the
dynamic exponent are significantly lower than the corresponding
Metropolis values.

\item In all cases, the magnetization is the observable that suffers
most from critical slowing down.  This behavior is quite different
from that of the Ising model on a regular lattice, where the energy
exhibits CSD equal to or greater than that of the magnetization
\cite{PAUL}.  In our simulations the observable $\vert q - 6 \vert$
did not show any significant CSD.

\begin{figure}[th]
\epsfxsize=\figsize \epsfbox{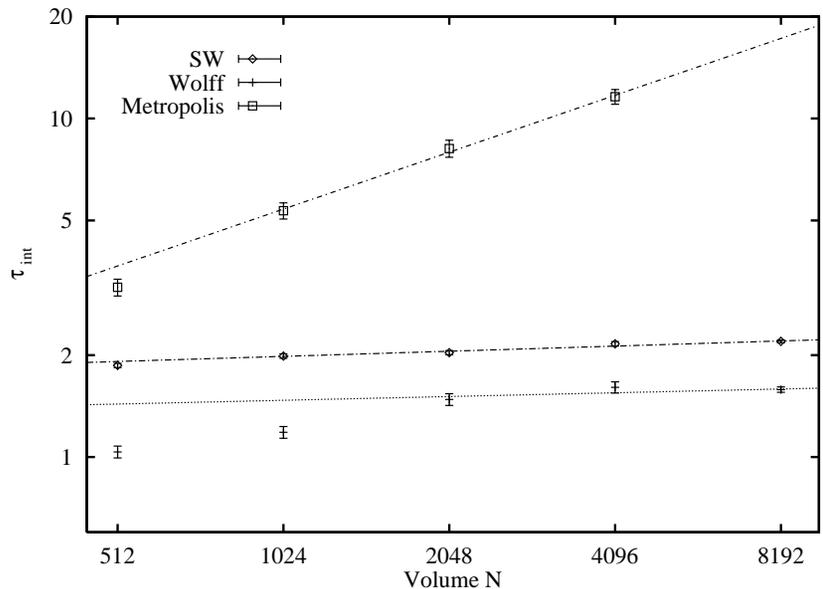}
\protect\caption{\protect\label{fig:1genene} Comparison of the integrated
autocorrelation times for the energy in the $n_{\mathrm{s}}=1$ model.}
\end{figure}

\item The SW and Wolff algorithms have similar performance within the
statistical accuracy of our data.  On the smaller lattices the Wolff
algorithm is somewhat more efficient than SW, but this advantage is a
finite size artifact since it disappears on the larger lattices.  For
\mbox{2--d} Ising models on flat and Poissonian lattices these algorithms
exhibit roughly comparable performance \cite{JANKE,PAUL}.  In the
\mbox{3--d} case, the Wolff algorithm is much more efficient.  It
seems that the relative performance of SW and Wolff algorithms depends
on the dimensionality of the lattice.

\begin{figure}[tbh]
\epsfxsize=\figsize \epsfbox{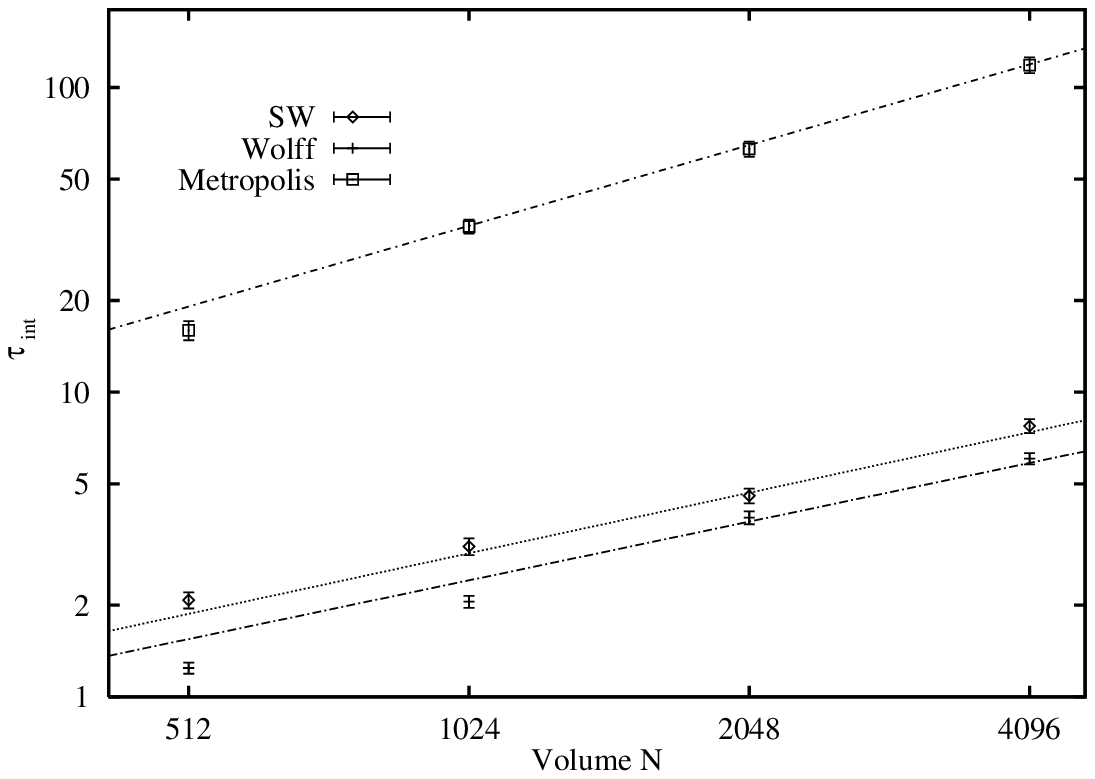}
\protect\caption{\protect\label{fig:2genmag} Comparison of the integrated
autocorrelation times for the magnetization in the $n_{\mathrm{s}}=2$
model.  We excluded smaller volume points from the fits.}
\end{figure}

\item \label{arg}
It is hard to determine differences in the degree of CSD between one
and two Ising models coupled to gravity, given our statistics.  In
\cite{TWOISING} we found, in fact, that the numerically measured
behavior of the one and two species models is very similar.  This,
however, is a consequence of logarithmic corrections \cite{KLEB} in
the two-species model.  One might suspect that the scaling law for
$\tau$ in this model incorporates logarithms as well.  On much larger
lattices where logarithmic behavior is distinguishable from small
power law scaling, the effective CSD might be considerably different
in the two generation, as compared to one generation, case.  This
situation is similar to that of the \mbox{2--d} 4--state Potts model.
Here, Li and Sokal \cite{LISOK} have shown that measurements of $z$,
obtained using the ansatz (\ref{ansatz}), violate rigorous bounds.
They suggest that these measurements of $z$ are not correct because
the fits to $\tau$ fail to take into account logarithmic corrections.
For the $n_{\mathrm{s}}=3$ model the situation is worse; the
corrections to scaling of (\ref{ansatz}) may be even larger.  In this
case, there are no theoretical arguments that predict the form of
these corrections.  Therefore, we anticipate that the numbers we have
quoted for $z/d_H$ differ considerably from the correct asymptotic
values in the $n_{\mathrm{s}}=2$ and $3$ cases.  To give a sense of
the magnitude of the corrections to scaling we note that on similar
size lattices, the estimate of $\gamma/\nu d_H$ differs by about 3\%
from its asymptotic value in the $n_{\mathrm{s}}=1$ case and by almost
50\% in the $n_{\mathrm{s}}=2$ case \cite{TWOISING}.
\end{enumerate}

\begin{figure}[t]
\epsfxsize=\figsize \epsfbox{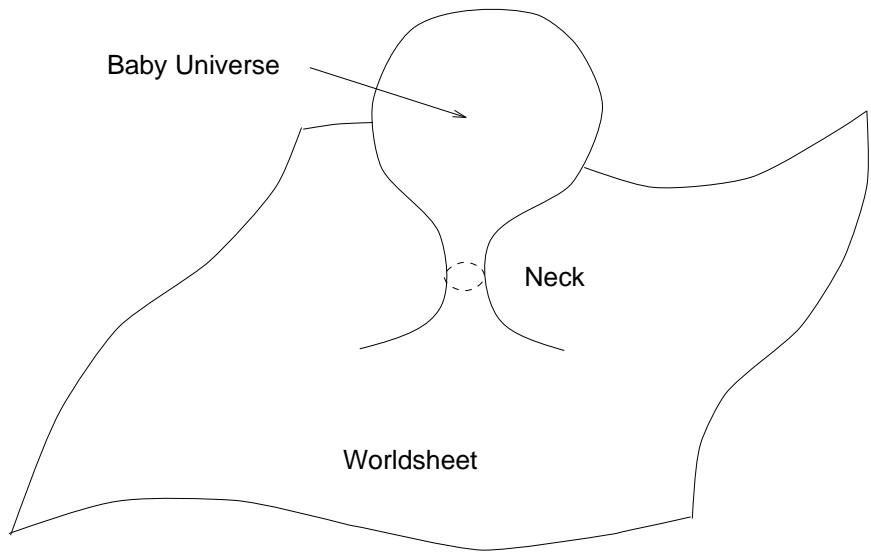}
\protect\caption{\protect\label{fig:baby} Schematic representation of
a baby universe.}
\end{figure}

The presence of CSD in these models should not be a surprise, since
the triangulation is updated locally.  In this context, it is relevant
to briefly recall the results of a similar analysis \cite{GEOFF} in
which percolation clusters were studied on random triangulations
without matter (pure gravity). The meshes were updated locally; all
link flips that did not lead to degenerate triangulations were
allowed.  It has been shown that these random triangulations are
characterized by a scaling distribution of baby universes \cite{JAIN}.
The formation of clusters is quite sensitive to the presence of the
bottlenecks (see fig.~\ref{fig:baby}), which inhibit cluster growth
into and out of baby universes.  Since this structure of baby
universes is slow to decorrelate under the local link-flip updates, we
expect that the mean percolation cluster sizes will be afflicted by
critical slowing down.  In the case of pure gravity, the mean size of
percolation clusters built on these triangulations in fact exhibits
critical slowing down of magnitude $z/d_H = .70(2)$.  We also
constructed percolation clusters in the Ising simulations that used SW
spin updates.  We observed, in this case, a similar CSD of the
world-sheet geometry; $z/d_H$ for the mean size of pure percolation
clusters was measured to be $.74(6)$.

We now argue that the critical slowing down in the gravity sector
should lead to considerable CSD for magnetic observables.  For one
would expect that baby universes should trap FK clusters as well as
pure percolation clusters.  As said before, for an Ising model
simulated on an arbitrary random triangulation, the mean FK cluster
size equals the average magnetic susceptibility.  The value of the
magnetization is thus clearly sensitive to those features of the
geometry that strongly affect the FK cluster size.  This coupling
transfers critical slowing down to the magnetic sector.  Some evidence
in support of this argument follows from our measurements of the mean
size of the FK clusters built to perform SW updates.  This observable
exhibited a value of $z/d_H$ of .52(6).

The efficiency of cluster algorithms is also typically affected by the
distribution of cluster sizes.  If the clusters are too small,
flipping them will fail to decorrelate distant spins.  If one cluster
fills most of the lattice, successive flips will essentially cancel
each other out.  Indeed FK clusters, on average, are much smaller in
the dynamical case than in the case of a fixed flat lattice.  Their
mean size is determined by the exponent $\gamma/\nu d_H$;
${\cal{S}}_{\rm{FK}} \sim N^{\gamma/\nu d_H}$.  For flat lattices,
$\gamma / \nu d_H = 7/8$. It is $2/3$ for the $n_{\mathrm{s}}=1$ Ising
model, which is quite close to the value for the 3--$d$ Ising
model\footnote{ Simulations of the \mbox{3--d} Ising model with
cluster algorithms suffer noticeable critical slowing down, though
still rather modest compared to the CSD observed here. In the flat
\mbox{3--d} case the energy is the observable showing the greater CSD.}.

It is difficult to determine directly whether the autocorrelation
times are influenced primarily by the slow decorrelation of the
world-sheet geometry or the effects of the smaller cluster sizes. The
argument, however, that CSD arises from updates of the geometry
applies only to magnetic observables, and not to the energy density.
Since the CSD we observe is greater for the magnetization, it seems
likely that the slow decorrelation of the world-sheet geometry is the
primary mechanism responsible for CSD.

\section*{Acknowledgments}
This work has been done with NPAC (Northeast Parallel Architectures
Center) and CASE computing facilities.  We would like to thank John
Apostolakis, Simon Catterall and Paul Coddington for helpful
correspondence and conversations.  MF wishes to thank G. Thorleifsson
for stimulating discussions.  The research of MB was supported by the
Department of Energy Outstanding Junior Investigator Grant DOE
DE--FG02--85ER40231, that of MF by funds from NPAC and that of GH by
research funds from Syracuse University.


\begin{thebibliography}{9}
\bibitem{SW}
R. H. Swendsen and J.-S. Wang, Phys. Rev. Lett.  58 (1987) 86.
\bibitem{WOLFF}
U. Wolff, Phys. Rev. Lett.  62 (1989) 361.
\bibitem{METRO}
N. Metropolis, A. W. Rosenbluth, M. N. Rosenbluth, A. H. Teller and E.
Teller, J. of Chem. Phys.  21 (1953) 1087.
\bibitem{REVIEWS}
A. D. Sokal, Computer Simulation Studies in Condensed Matter Physics:
Recent Developments, eds. D. P. Landau, K. K. Mon and H. -B.
Sch\"{u}ttler (Springer-Verlag, Berlin-Heidelberg, 1988).\\ U.  Wolff,
Proc. of the 1989 Symposium on Lattice Field Theory, Capri, Nucl.
Phys.  B (Proc. Suppl.)  17 (1990) 93.
\bibitem{DTRS1}
C. Baillie and D. Johnston, Mod. Phys. Lett.  A7 (1992) 1519; Phys.
Lett.  B286 (1992) 277.
\bibitem{DTRS2}
S. M. Catterall, J. B. Kogut and R. L. Renken, Phys. Rev.  D45 (1992)
2957; Phys. Lett.  B292 (1992) 277.
\bibitem{DTRS3}
J Ambj{\o}rn, B. Durhuus, T. J\'onsson and G. Thorleifsson, Nucl.
Phys.  B398 (1993) 568.
\bibitem{ANALYTICAL}
V.A. Kazakov, Phys. Lett.  A119 (1986) 140.\\ D. V. Boulatov and V. A.
Kazakov, Phys. Lett.  B186 (1987) 379.
\bibitem{TWOISING}
M. Bowick, M. Falcioni, G. Harris and E. Marinari, Two Ising Models
Coupled to 2--Dimensional Gravity, Syracuse Univ.  preprint
SU-HEP-93--4241--556 and hep-th/9310136.
\bibitem{GEOFF}
G. Harris, Percolation on Strings and the Cover-up of the c = 1
Disaster, Syracuse Univ. preprint SU-HEP-93--4241--555 and
hep-th/9310137.
\bibitem{FLIP}
D. Boulatov, V. Kazakov, I. Kostov and A. A. Migdal. Phys. Lett.  B157
(1985) 295.
\bibitem{FK}
C. M. Fortuin and P. W. Kasteleyn, Physica 57 (1972) 536.
\bibitem{WOLFF2}
U. Wolff, Phys. Lett.  B228 (1989) 379.
\bibitem{PERCOLBOOK}
D. Stauffer and A. Aharony, Introduction to Percolation Theory, Taylor
and Francis, London, U.K., 1992.
\bibitem{SWEENY}
M. Sweeny, Phys. Rev.  B27 (1983) 4445.
\bibitem{SOKAL}
A. D. Sokal, Monte Carlo Methods in Statistical Mechanics: Foundations
and Algorithms, NYU preprint based on the lectures at the Troisi\`eme
Cycle de la Physique en Suisse Romande, June 1989.
\bibitem{JANKE}
W. Janke, M. Katoot, and R. Villanova, preprint UAB-FT-318 and
hep-lat/9310025.
\bibitem{BINDER}
K. Binder, Z. Phys.  B43 (1981) 119.
\bibitem{PAUL}
C.~F.~Baillie and P.~D.~Coddington, Phys. Rev. B43 (1991) 10617.
\bibitem{KLEB}
I. Klebanov, String Theory in Two-Dimensions, in Proc.  of the Trieste
School on String Theory and Quantum Gravity '91, hep-th/9108019.
\bibitem{LISOK}
X.-J.~Li and A.~D.~Sokal, Phys. Rev. Lett.  63 (1989) 827.
\bibitem{JAIN}
S.~Jain and S.~Mathur, Phys. Lett.  B286 (1992) 239; J.~Ambj\o rn,
S.~Jain and G.~Thorleifsson, Phys. Lett.  B307 (1993) 34.
\end{thebibliography}
\end{document}